\documentclass[aps,pre,reprint,superscriptaddress,amsmath,amssymb,longbibliography]{revtex4-2}

\usepackage{booktabs}
\usepackage{mathtools}
\usepackage{bm}
\usepackage{graphicx}
\usepackage{hyperref}
\hypersetup{colorlinks=true, linkcolor=blue, citecolor=blue, urlcolor=blue}

\begin{document}

\title{A Geometric Foundation for the Universal Laws of Turbulence}

\author{Marcial Sanchis-Agudo}
\email{sanchis@kth.se}
\affiliation{FLOW, Engineering Mechanics, KTH Royal Institute of Technology, SE-100 44 Stockholm, Sweden}

\author{Ricardo Vinuesa}
\affiliation{Department of Aerospace Engineering, University of Michigan, Ann Arbor, MI 48109, USA}
\affiliation{FLOW, Engineering Mechanics, KTH Royal Institute of Technology, SE-100 44 Stockholm, Sweden}

\date{\today}

\begin{abstract}
We propose a theoretical framework where the dissipative structures of turbulence emerge from microscopic path uncertainty. By modeling fluid parcels as stochastic tracers governed by the Schr\"odinger Bridge (SB) variational principle, we demonstrate that the Navier--Stokes viscous term is a natural linear, second-order macroscopic operator consistent with isotropic microscopic diffusion. We derive two foundational pillars of turbulence from this single principle. First, we show that the Kolmogorov scale $\eta \sim (\nu^3/\epsilon)^{1/4}$ is not merely a dimensional necessity but a geometric diffusion horizon: it is the scale at which the kinetic energy of a fractal trajectory, scaling as $k \sim \nu/\tau$, balances the macroscopic dissipation rate. Second, we show that the universal law of the wall is the stationary solution to this stochastic process under no-slip constraints. The logarithmic mean profile arises from the scale invariance of the turbulent diffusivity, while finite-Reynolds-number corrections emerge as controlled asymptotic expansions of the stochastic variance. This framework offers a physically grounded derivation of turbulent scaling laws that complements and extends purely phenomenological dimensional analysis.
\end{abstract}

\maketitle

\section{Introduction}

Turbulence is governed by two celebrated scaling laws: the Kolmogorov scale for dissipation in the bulk \cite{Kolmogorov1941}, and the logarithmic law of the wall near a solid boundary \cite{vonKarman1930}. Despite their centrality to fluid dynamics, these laws have generally been derived from dimensional analysis and phenomenological arguments \cite{frisch1995turbulence}. Moreover, the universality of the law of the wall remains controversial: meticulous data analyses show the von K\'arm\'an constant $\kappa$ varies with geometry \cite{Nagib2008}, while asymptotic theories suggest universality \cite{Luchini2017PRL,Monkewitz2017PRF}.

In this work, we propose a change in perspective: we treat the fluid not as a continuum of determinate paths, but as an ensemble of stochastic trajectories. 

\begin{figure}[b]
\includegraphics[width=0.95\linewidth]{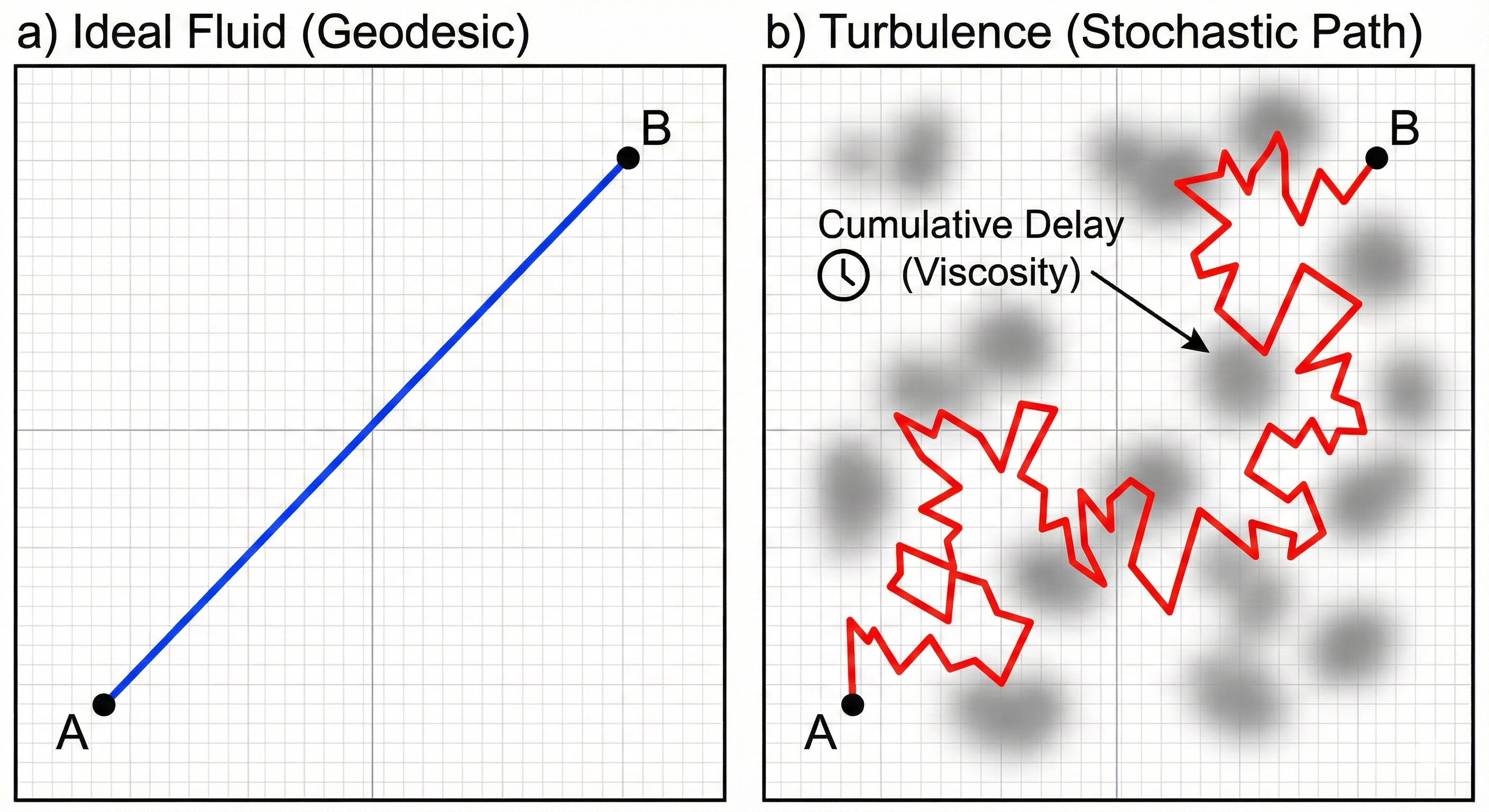} 
\caption{\label{fig:plaza} The geometric origin of viscosity. (a) In an ideal fluid, parcels follow geodesics (paths of least action). (b) In a turbulent fluid, microscopic path uncertainty forces a stochastic trajectory. The macroscopic slowness relative to the geodesic time defines the effective viscosity, and the entropic cost of this deviation is the dissipation.}
\end{figure}

\emph{Thought experiment.} Imagine crossing a grand plaza from A to B. If the plaza is empty, your path is a straight line (a geodesic), the world of least action, as in Arnold's description of ideal fluids \cite{Arnold1966,SanchisAgudo2025PoF}. If the plaza is crowded, your intent remains, but your trajectory is stochastically perturbed: you navigate a \emph{cloud of paths}. The intensity of this exploration is the effective diffusivity $\nu_{\text{eff}} = \nu + s(\mathbf{x})$, where $\nu$ is the baseline molecular tremor and $s$ is the emergent turbulent jostling. The cumulative delay, measured relative to geodesic time, defines an emergent \emph{slowness} that we identify with viscosity, while the chaotic jostling corresponds to dissipation. This transition from ideal geodesic motion to stochastic diffusion is illustrated in Fig.~\ref{fig:plaza}.

This thought experiment forms the foundation of our theory, where we posit that viscous dissipation is the macroscopic signature of this microscopic path uncertainty. This single principle has two major, independent consequences that together structure turbulent flows.

First, for the bulk flow, it replaces Kolmogorov's phenomenology with two falsifiable physical laws rooted in diffusion physics: (i) an energy-balance law, which dictates that the macroscopic turbulent-kinetic-energy (TKE) dissipation rate $\epsilon$ must equal the rate required to sustain the microscopic stochastic tremor; and (ii) a diffusion-length law, which defines the dissipation scale $\eta$ as the characteristic distance the process diffuses over its lifetime.

Second, near a boundary, the same principle dictates that the streamwise velocity profile $U(y)$ must be a stationary state of the constrained stochastic process. This constraint mathematically enforces a local balance equation, $(sU')' = 0$, where $s(y)$ is the intensity (diffusivity) of the stochastic tremor. The universal logarithmic law then emerges as a direct consequence of the linear scaling of this diffusivity with distance from the wall, $s(y) \propto y$.

\section{The Variational Principle: \\ Viscosity from Uncertainty}
\label{sec:variational}

The Navier--Stokes equations synthesize conservative and dissipative dynamics \cite{landau1987fluid}. The conservative part is described geometrically by Arnold's geodesic framework for the Euler equations \cite{Arnold1966}. We identify the dissipative term with an entropic origin: uncertainty in the space of microscopic fluid paths.

\subsection{Microscopic Stochastic Model}
We model the motion of a microscopic fluid parcel $\mathbf{X}_t$ as a diffusion process that accounts for both molecular and turbulent tremor. The trajectory is governed by the Stochastic Differential Equation (SDE):
\begin{equation}
\mathrm{d}\mathbf{X}_t 
  = \mathbf{u}(\mathbf{X}_t, t)\,\mathrm{d}t 
  + \sqrt{2\big(\nu + s(\mathbf{X}_t,t)\big)}\,\mathrm{d}\mathbf{W}_t,
\end{equation}
where $\mathbf{u}$ is the coarse-grained drift (later identified with the Schr\"odinger-Bridge current velocity and with the macroscopic flow), $\nu$ is the kinematic viscosity, $s(\mathbf{x},t)$ is the emergent stochastic diffusivity representing the turbulence intensity, and $\mathbf{W}_t$ is a standard Wiener process. This formulation explicitly introduces the variable diffusivity $\nu+s(\mathbf{x},t)$ at the microscopic level.

\subsection{Schr\"odinger Bridge and macroscopic dissipation}

To extract the macroscopic form of the resulting force, we use the Schr\"odinger Bridge (SB) problem \cite{schrodinger1931umkehrung,leonard2014survey}. The SB problem seeks the most likely evolution of the probability density $\pi(\bm{x},t)$ connecting two states, minimizing an action functional that penalizes deviations from a reference Wiener measure.

For a diffusion with scalar, isotropic diffusivity 
$D(\bm{x}) = \nu + s(\bm{x})$ (so that over a short time step $\Delta t$ the kernel variance satisfies $\langle|\Delta\mathbf{X}|^2\rangle \sim 2D\,\Delta t$), the associated Fokker--Planck equation can be written in conservative form
\begin{equation}
\partial_t \pi + \nabla\!\cdot(\pi \,\bm{u}) 
  = \nabla \cdot \big( D \nabla \pi \big),
\label{eq:FP}
\end{equation}
where $\bm{u}$ is the SB \emph{current velocity} (the drift corrected by the osmotic term, see e.g.\ \cite{Holm2015,Koide2012}). In what follows we identify this SB current velocity with the macroscopic velocity field: $\bm{u}\equiv\bm{v}$ at the level of resolved dynamics.

In the stochastic variational formulations of fluid dynamics \cite{Holm2015,Koide2012}, the SB action with such a constraint leads to an effective momentum balance for the macroscopic velocity $\bm{v}$ of the form
\begin{equation}
\rho\big(\partial_t \bm{v} + (\bm{v}\!\cdot\!\nabla)\bm{v}\big)
  = -\nabla p 
  + \nabla \cdot \big(\rho\, D \nabla \bm{v}\big)
  + \text{(gradient terms)},
\end{equation}
where $\rho$ is the fluid mass density. For an incompressible flow with slowly varying scalar diffusivity $D(\bm{x}) = \nu + s(\bm{x})$, the additional gradient terms can be absorbed into a redefinition of the pressure, leaving the structural form
\begin{equation}
\rho\big(\partial_t \bm{v} + (\bm{v}\!\cdot\!\nabla)\bm{v}\big)
  = -\nabla p 
  + \nabla \cdot \big(\rho\,(\nu+s)\,\nabla \bm{v}\big),
\label{eq:NS_structural}
\end{equation}
which we interpret as the \emph{entropic force} induced by microscopic path uncertainty.

We emphasize that we do not postulate this form ad hoc. The divergence structure $\nabla \cdot (D \nabla \cdot)$ emerges from the Schr\"odinger-Bridge variational problem under the assumption of scalar, isotropic microscopic diffusivity $\nu+s(\mathbf{x})$. Within this modeling class, it is the natural linear, second-order, objective operator compatible with the Fokker--Planck constraint, up to pressure redefinitions. The subsequent determination of the profile $s(y)$ in wall-bounded flows relies on additional symmetry, constant-stress and scale-separation arguments, which we make explicit below.

\subsection*{Assumptions and scope}

Before turning to the consequences for turbulence, it is useful to summarize the assumptions underlying our framework. At the microscopic level we assume that fluid parcels follow diffusion processes with scalar, isotropic diffusivity $\nu + s(\mathbf{x})$; anisotropic effects at very high Reynolds number are assumed to be captured effectively through the spatial dependence of $s(\mathbf{x})$. Their collective statistics are governed by a Schr\"odinger-Bridge variational principle. At the macroscopic level we adopt a mean-field closure in which the drift velocity $\mathbf{u}$ coincides with the resolved mean flow $\mathbf{v}$, and in wall-bounded shear flows we exploit the classical separation between inner and outer scales. In the overlap region we further assume scale invariance of the one-dimensional operator $\mathcal{L} = \partial_y\bigl(s(y)\partial_y\bigr)$ under wall-normal dilations. The results that follow, the Kolmogorov scale as a diffusion horizon and the logarithmic (plus linear) law of the wall, should therefore be interpreted as consequences of these explicit hypotheses, rather than as theorems valid for arbitrary dissipative flows.

\section{First Consequence: The Kolmogorov Scale}

We now apply this stochastic framework to the smallest scales of turbulence. Instead of assuming dimensional homogeneity, we treat the dissipation scale as the limit where the kinetic energy of the stochastic fluctuations is fully thermalized.

\subsection{Energy Balance of a Fractal Path}

In a standard ballistic (Eulerian) trajectory, velocity is independent of the observation timescale. However, the trajectory defined by the SDE is fractal. Per Einstein--Smoluchowski \cite{Einstein1905}, the mean-squared displacement over a time $\tau$ scales as
\begin{equation}
\langle |\Delta\mathbf{x}|^2 \rangle_\tau \;=\; C_d\,\nu\,\tau,
\end{equation}
where $C_d=2d$ for isotropic Brownian motion. Note that at the smallest scales, the turbulent contribution $s$ decays to zero, so that the total diffusivity $\nu_{\text{eff}}=\nu + s$ approaches the molecular floor $\nu$, and the scaling is controlled by molecular viscosity.

This implies that the characteristic velocity fluctuation $v_\tau$ observed over a timescale $\tau$ scales as:
\begin{equation}
v_\tau \sim \frac{\sqrt{\langle |\Delta\mathbf{x}|^2 \rangle_\tau}}{\tau} \sim \frac{\sqrt{\nu\tau}}{\tau} = \sqrt{\frac{\nu}{\tau}}.
\end{equation}
The specific kinetic energy $k_\tau$ associated with this stochastic tremor is therefore scale-dependent:
\begin{equation}
k_\tau \sim v_\tau^2 \sim \frac{\nu}{\tau}.
\end{equation}
The rate at which this energy is processed (the stochastic power) is the energy divided by the timescale:
\begin{equation}
P_{\text{stoch}}(\tau) \sim \frac{k_\tau}{\tau} \sim \frac{\nu}{\tau^2}.
\end{equation}

\subsection{Scale Selection and the Kolmogorov Time}
The turbulent cascade transfers energy from large scales at a constant macroscopic rate $\epsilon$. This cascade proceeds down to smaller scales (decreasing $\tau$) until the kinetic energy capacity of the stochastic tremor rises to meet the supply. The Kolmogorov timescale $\tau_\eta$ is defined as the unique scale where the stochastic processing rate matches the macroscopic input:
\begin{equation}
P_{\text{stoch}}(\tau_\eta) \equiv \epsilon \quad\Rightarrow\quad \epsilon = \frac{\nu}{\tau_\eta^2}.
\end{equation}
Solving for $\tau_\eta$, we obtain the characteristic time scale:
\begin{equation}
\tau_\eta = \left(\frac{\nu}{\epsilon}\right)^{1/2}.
\label{eq:tau_eta_derivation}
\end{equation}

\subsection{The Diffusion-Length Law}
The Kolmogorov length scale $\eta$ is traditionally found by combining $\nu$ and $\epsilon$. In our framework, $\eta$ is a geometric consequence: it is the characteristic distance the stochastic tremor diffuses during its lifetime $\tau_\eta$. Using the diffusion law $\eta^2 \sim \nu \tau_\eta$ and substituting Eq.~\eqref{eq:tau_eta_derivation}:
\begin{equation}
\eta \sim \sqrt{\nu \left(\frac{\nu}{\epsilon}\right)^{1/2}} = \left(\frac{\nu^3}{\epsilon}\right)^{1/4}.
\end{equation}
Thus, the Kolmogorov scale emerges from the kinematics of diffusion and energy conservation, without invoking the Buckingham $\pi$ theorem.

\section{Second Consequence: The Universal Law of the Wall}

We now consider the flow bounded by a solid wall at $y=0$. The no-slip condition imposes a constraint on the stochastic process. The mean velocity profile $U(y)$ must be a stationary state (a fixed point) of the stochastic evolution near the boundary.

This stationarity is governed by an integral equation with the \emph{Dirichlet} heat kernel $K_D$ \cite{Carslaw1959,Redner2001}:
\begin{align}
U(y)&=\int_{0}^{\infty} K_D\!\left(y,y';\,\nu+s(y)\right)\,U(y')\,\mathrm{d}y',
\label{eq:fp}
\end{align}
where the kernel is generated by the total diffusivity $\nu+s(y)$. A controlled short-time parametrix expansion of \eqref{eq:fp} \cite{Friedman1964,Olver1974,Gilkey1995} recovers the generator balance:
\begin{equation}
\frac{\mathrm{d}}{\mathrm{d}y}\left( (\nu+s(y)) \frac{\mathrm{d}U}{\mathrm{d}y} \right) = 0.
\label{eq:kernel-balance}
\end{equation}
In the overlap region, the turbulent diffusivity dominates ($s \gg \nu$), simplifying the balance to $(sU')'=0$. This implies $s(y)U'(y) \approx u_\tau^2$, identifying $s(y)$ as the effective viscosity connecting mean strain to wall stress.

\begin{figure}[t]
\includegraphics[width=0.95\linewidth]{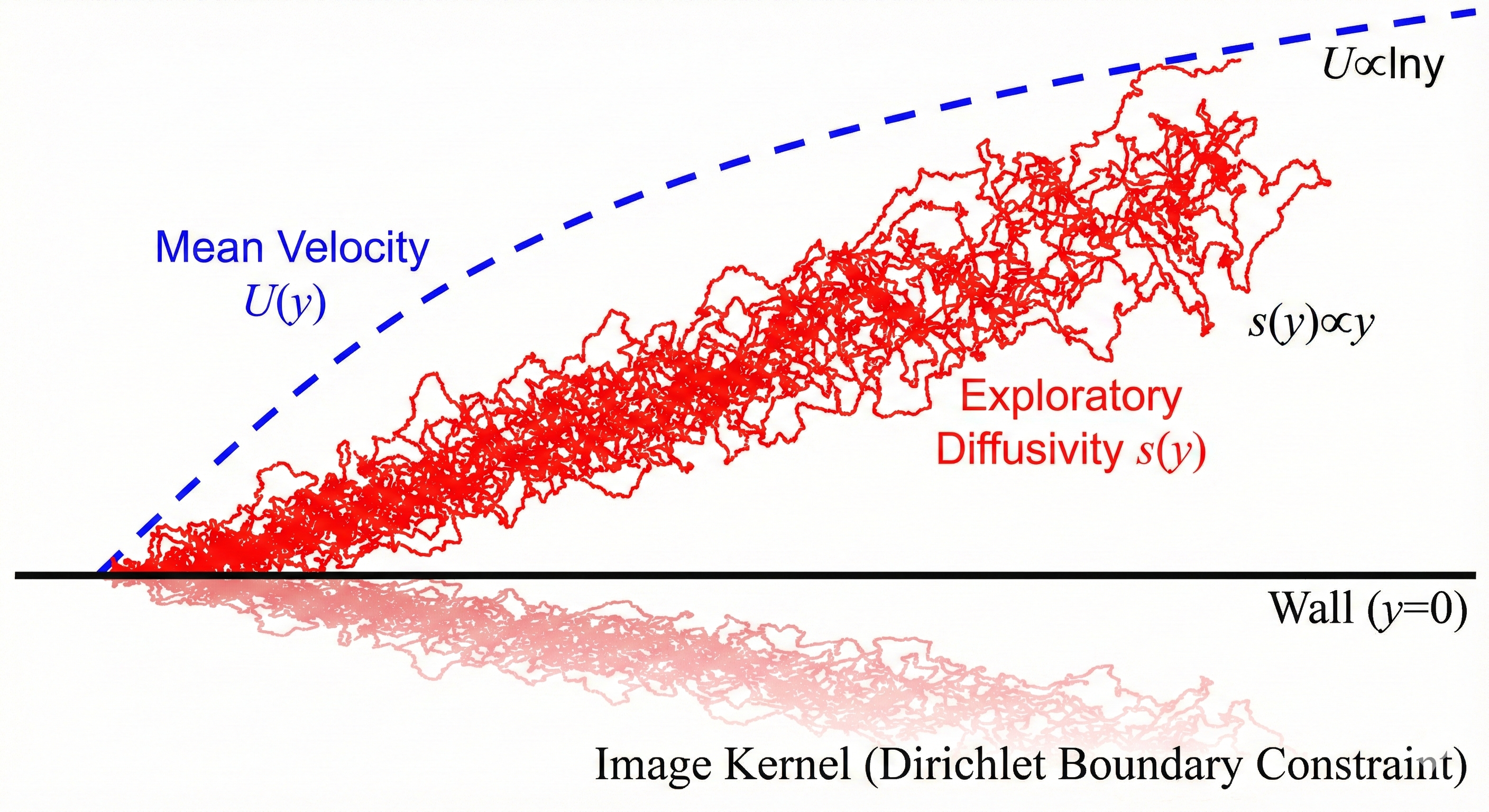} 
\caption{\label{fig:wall_law} The Law of the Wall as a geometric fixed point. The stochastic diffusivity (red paths) scales linearly with wall distance, $s(y)\propto y$, due to scale invariance. The mean velocity profile (blue dashed line) that satisfies the stationarity condition for this expanding variance is necessarily logarithmic, $U \propto \ln y$. The faded reflection illustrates the Dirichlet image kernel enforcing the no-slip condition.}
\end{figure}

\subsection{Scale Invariance and the Logarithmic Law}
In the overlap region (inertial sublayer), the physics implies scale invariance of the operator $\mathcal{L}=\partial_y(s\partial_y)$ under the transformation $y \mapsto \lambda y$. This symmetry naturally singles out a linear scaling of the stochastic diffusivity:
\begin{equation}
s(y) = c\,y,
\end{equation}
where $c$ is a velocity scale. This linear expansion of path uncertainty is illustrated in Fig.~\ref{fig:wall_law}. Substituting this into the momentum balance $s(y)U'(y) = u_\tau^2$ identifies the constant $c = \kappa u_\tau$, yielding:
\begin{equation}
U'(y) = \frac{u_\tau}{\kappa y} \quad \implies \quad U^+(y^+) = \frac{1}{\kappa} \ln y^+ + B,
\end{equation}
where $U^+ = U/u_\tau$ and $y^+ = y u_\tau / \nu$ are the usual inner-scaled variables. Here, the von K\'arm\'an constant $\kappa$ is interpreted not as an empirical mixing coefficient, but as the growth rate of the stochastic diffusivity $s(y)$.

\subsection{Asymptotic Corrections}
Finite–$Re$ outer effects (nonuniform shear, mild curvature, pressure gradient) appear as a small $O(y/\delta)$ departure from strict linearity in the diffusivity:
\begin{equation}
s(y)=\kappa u_\tau y\Big(1+a\,\frac{y}{\delta}+O((y/\delta)^2)\Big).
\end{equation}
Substituting this into the balance equation yields the extended overlap law:
\begin{equation}
U^+(y^+) \approx \frac{1}{\kappa} \ln y^+ + B + S_0 \frac{y}{\delta},
\end{equation}
where $S_0 = -a/\kappa$. This linear correction term, often observed in high-fidelity DNS data as a wake or pressure-gradient modification, emerges here as a higher-order term in the expansion of the stochastic variance. This suggests that universality holds asymptotically, while geometry-dependent deviations at finite $Re$ are structural consequences of the bounded diffusion domain.

\section{Theoretical Interpretation of the Universality Debate}

The full kernel balance is geometry-dependent because the Laplacian differs by coordinates. For a pipe, cylindrical geometry introduces a metric-curvature term:
\begin{equation}
\nabla \cdot \big( (\nu+s) \nabla U \big) = \frac{1}{r}\frac{\mathrm{d}}{\mathrm{d}r}\!\left(r\,(\nu+s)\,U_r\right)=0.
\end{equation}
Assuming $s \gg \nu$ in the core and using the wall-normal coordinate $y=R-r$ (so $r=R-y$ and $U_r = -U'$), we obtain the balance:
\begin{equation}
\frac{\mathrm{d}}{\mathrm{d}y}\!\left[(R-y)\,s(y)\,U'(y)\right]=0.
\end{equation}
Expanding the derivative product yields:
\begin{equation}
(sU')' - \frac{s}{R-y}U' = 0.
\label{eq:pipe-balance}
\end{equation}
This metric-curvature term vanishes as $y/R\to 0$, explaining asymptotic universality. However, it is finite at any fixed Reynolds number and is the origin of the stronger wake observed in pipes compared to channels. In the language of the extended overlap, Eq.~\eqref{eq:pipe-balance} adds an $O(y/R)$ correction to the effective variance, equivalent to $s(y)=c\,y\,[1+O(y/R)]$. This generates a positive linear correction in $U$ (a larger apparent $\kappa$ at finite $Re$) that decays as the overlap widens with $Re_\tau$~\cite{Luchini2017PRL,Monkewitz2017PRF,Hoyas2024PRF}. Pressure gradients act analogously through a nonuniform $\tau^+(y)$, producing $O(y/\delta)$ departures from $s\propto y$ and thus a measurable $S_0$ in the extended overlap. This reconciles asymptotic universality with finite-$Re$ variability in $\kappa$ across geometries and PG conditions~\cite{Nagib2008}. This interpretation is consistent with geometric viewpoints where pressure and wall stresses are encoded in curvature-dependent boundary actions and their associated stress–energy tensors \cite{SanchisAgudo2025PoF}.

\section{Conclusions}
Returning to our analogy of the grand plaza, we have moved beyond metaphors to provide mathematical and physical substance. We proposed a geometric foundation for turbulence where viscous dissipation emerges from uncertainty in the space of microscopic fluid paths. This single principle has led to two significant, verifiable consequences derived from a single geometric closure and explicit physical hypotheses. First, the Kolmogorov scale is not merely a result of dimensional analysis but follows from two physical laws: (1) an energy balance, stating that the TKE dissipation rate $\epsilon$ equals the rate at which the microscopic tremor processes kinetic energy, $\nu/\tau_\eta^2$, which defines the Kolmogorov time $\tau_\eta$; and (2) a diffusion principle, stating that the Kolmogorov length $\eta$ is the distance the tremor diffuses in that time, $\eta^2 \sim \nu \tau_\eta$. This yields the celebrated scaling $\eta \sim (\nu^3/\epsilon)^{1/4}$ from a self-contained physical argument anchored by the Einstein--Smoluchowski diffusion law \cite{Einstein1905,Smoluchowski1906,Reif2009}. Second, the universal law of the wall is not an empirical fit but the unique velocity profile a collective must adopt to be in a stationary state with its own exploratory dynamics near a boundary. The structural equivalence, that $U$ is logarithmic if and only if $s$ is linear within this class of models, serves as a concrete bridge between the macroscopic world we measure and the microscopic world of path uncertainty we postulate. The observed \emph{log+linear} overlap with a linear coefficient that vanishes as $Re_\tau\!\to\!\infty$~\cite{Hoyas2024PRF} is a direct, independent validation of the controlled departures $s(y)=cy(1+a\,y/\delta)$ predicted by our framework, and clarifies the von K\'arm\'an debate by attributing finite-$Re$ variations to geometry- and PG-induced $O(y/\delta)$ corrections to this fundamental linear scaling of $s(y)$.

The present framework should thus be viewed as a consistent, physically motivated closure that unifies bulk and wall scaling under a single geometric principle of path uncertainty. It does not attempt to solve turbulence in full generality, but rather to illuminate why the classical Kolmogorov and von Kármán laws are so robust, and how their finite-$Re$ deviations can be interpreted in terms of the geometry and constraints of the underlying diffusion process. Together with recent geometric formulations of pressure and wall stresses in inviscid potential flows~\cite{SanchisAgudo2025PoF}, the present work forms part of a broader program to recast classical turbulent and wall-bounded phenomena in terms of variational principles and geometric structures.

\begin{acknowledgments}
M.S.-A.\ acknowledges financial support from the
EU Doctoral Network MODELAIR. M.S.-A.\ thanks Fermin Mallor Franco, Francisco-Javier Granados-Ortiz, Miguel P.\ Encinar, Ozan \"Oktem and Emelie Saga Stark for the insightful conversations and enlightening perspectives. Special thanks to Francesco Mario D'Afiero and Eduardo Terres-Caballero for being both receptive and scientifically critical. Finally, M.S.-A.\ thanks Ricardo Vinuesa for his support, supervision and trust in this research.
\end{acknowledgments}

\bibliography{pre}
\end{document}